\documentclass[%
 aip,
 cha,
 amsmath,amssymb,
 reprint,%
]{revtex4-1}

\usepackage{graphicx}
\usepackage{dcolumn}
\usepackage{bm}

\usepackage[utf8]{inputenc}
\usepackage[T1]{fontenc}
\usepackage{mathptmx}
\usepackage{etoolbox}

\makeatletter
\def\@email#1#2{%
 \endgroup
 \patchcmd{\titleblock@produce}
  {\frontmatter@RRAPformat}
  {\frontmatter@RRAPformat{\produce@RRAP{*#1\href{mailto:#2}{#2}}}\frontmatter@RRAPformat}
  {}{}
}%
\makeatother
\def\eps{{\varepsilon}}
\begin{document}

\preprint{AIP/123-QED}

\title[]{Waveform proportionality and Taylor's law in networks of Lorenz systems}
\author{Yuzuru Mitsui}
\email{mitsui@design.kyushu-u.ac.jp}
 \altaffiliation[Also at ]{Education and Research Center for Mathematical and Data Science, Kyushu University, Motooka, Fukuoka, 819-0395, Japan.}
 \affiliation{%
 Faculty of Design, Kyushu University, Shiobaru, Fukuoka, 815-8540, Japan
}

\author{Hiroshi Kori}%
\affiliation{ 
Department of Complexity Science and Engineering, The University of Tokyo, Kashiwa, Chiba 277-8561, Japan
}%

\date{\today}

\begin{abstract}
Taylor's law (TL), a power-law relationship between the mean and variance of a quantity, has been observed across diverse scientific disciplines. Despite its ubiquity, the underlying mechanisms responsible for TL have not yet been fully elucidated. In particular, the frequent empirical observation of TL with an exponent 2 warrants further investigation.  In a previous study, we hypothesized that synchronization contributes to the emergence of TL with an exponent 2.  To validate this hypothesis, we employed coupled oscillator models, with each oscillator described by a distinct dynamical system: a food chain model, the Rössler system, the Brusselator, and the Lorenz system. Our analytical and numerical results demonstrated that strong coupling leads to a form of synchronization wherein time series become proportional to each other, resulting in TL with an exponent 2. Here, we extend our previous findings for the coupled Lorenz system and provide detailed calculations. We analytically and numerically show that, under strong coupling, waveform proportionality and TL with an exponent 2 emerge for some variables in the globally coupled original, generalized, and hyperchaotic Lorenz systems. Furthermore, we derive a condition under which these properties hold for the original Lorenz system on networks. We then numerically confirm that when heterogeneous oscillators are randomly distributed within the network, waveform proportionality and TL with an exponent 2 emerge across several representative network structures, such as a two-dimensional lattice, random networks, and scale-free networks.
\end{abstract}

\maketitle

\begin{quotation}
Taylor's law (TL) states that the variance of a quantity, such as population abundance, scales as a power of its mean. Theoretical studies have shown how such scaling can arise, yet two questions remain: why is it so widespread in ecosystems, and why is the TL exponent so often close to two? Building on our previous work, we investigate synchronization as a possible mechanism for this scaling through analytical and numerical studies of coupled Lorenz systems. We show that, under appropriate coupling conditions, oscillators with different parameters can produce approximately proportional time series: their waveforms have nearly the same shape but different amplitudes. Consequently, the variance is proportional to the square of the mean, corresponding to TL with an exponent 2. This mechanism operates for both periodic and chaotic dynamics and extends to higher-dimensional Lorenz systems. We also derive a condition for this mechanism to operate on networks and numerically confirm the predicted scaling for several network structures when oscillators with different parameters are randomly assigned to nodes.
\end{quotation}

\section{Introduction}
Taylor's law (TL)~\cite{Bliss_1941, Taylor_1961_Nature}, also known as fluctuation scaling, quantifies a power-law relationship between the mean and variance of a variable. This law has been extensively observed across various fields, with its origins in ecology~\cite{Eisler_etal_2008_AP, Taylor_2019_book}. TL is typically categorized into two types based on the method of calculating the mean and variance: temporal TL and spatial TL. Temporal TL utilizes the time average and variance of time series, whereas spatial TL employs the ensemble average and variance. Denoting the proportionality coefficient and exponent of temporal TL as $\alpha_{\textrm{t}}$ and $\beta_{\textrm{t}}$, respectively, and those of spatial TL as $\alpha_{\textrm{s}}$ and $\beta_{\textrm{s}}$, respectively, TL is expressed as
\begin{align}
    \log(\mathrm{variance}) = \log \alpha_{\mathrm{t,s}} + \beta_{\mathrm{t,s}} \times \log(\mathrm{mean}).
\end{align}
\noindent When the exponent takes values greater than one, this power-law relationship is often referred to as giant (number/density) fluctuations, a phenomenon extensively studied in the field of active matter physics~\cite{GNF_Ginelli_2016, GNF_Chate_2019, Nishiguchi_2023_JPSJ}.
Empirical data on population abundance in ecosystems frequently yield TL exponents close to 2 (Refs.~\onlinecite{Taylor_Woiwod_1982_JAE, Kerkhoff_Ballantyne_2003_EL, Zhao_etal_2019_JAE}). However, theoretical investigations have demonstrated that the TL exponent can take arbitrary real values~\cite{Cohen_etal_2013_PRSB, Cohen_2014_TE, Cohen_2014_TPB}, stimulating interest in understanding the mechanisms underlying the frequent emergence of TL with an exponent 2 (Ref.~\onlinecite{Giometto_etal_2015_PNAS}). In this context, we posited that synchronization underlies the emergence of TL with an exponent 2 (Ref.~\onlinecite{Mitsui_Kori_2025_PRL}), as several studies have indicated that stronger correlations between time series tend to drive the TL exponent towards 2 (Refs.~\onlinecite{Kerkhoff_Ballantyne_2003_EL, Ballantyne_Kerkhoff_2005_JTB, Hanski_1987}). Intriguingly, when the correlation becomes sufficiently strong such that the time series become proportional to each other, spatial TL with an exponent 2 emerges~\cite{Reuman_etal_2017_PNAS}. In Ref.~\onlinecite{Mitsui_Kori_2025_PRL}, we investigated our hypothesis using coupled dynamical system models and discovered that strong coupling induces a type of synchronization in which time series become mutually proportional, resulting in TL with an exponent 2. We refer to this type of synchronization as \textit{waveform proportionality} (WP)~\cite{Mitsui_Kori_2025_PRL}.

Although synchronization characterized by proportional time series has been previously reported as projective synchronization~\cite{Mainieri_Rehacek_1999_PRL} or as a specific instance of generalized synchronization~\cite{Kano_Umeno_2022_Chaos}, these observations were confined to particular classes of models. More recently, we demonstrated that this type of synchronization can manifest in a broader range of models~\cite{Mitsui_Kori_2025_PRL}. In our earlier work~\cite{Mitsui_Kori_2025_PRL}, we employed the coupled food chain model~\cite{Blasius_etal_1999_Nature, Blasius_Stone_2000_IJBC, Montbrio_Blasius_2003_Chaos}, coupled Rössler system~\cite{Rosenblum_etal_1996_PRL, Rosenblum_etal_1997_PRL, Sakaguchi_2000_PRE, Pikovsky_etal_1996_EPL,Montbrio_Blasius_2003_Chaos, Osipov_etal_1997_PRE}, coupled Brusselator~\cite{Daido_2011_PRE}, and coupled Lorenz system~\cite{Lee_etal_1998_PRL}. For the initial three models, we elucidated the mechanism of TL using a perturbative method and an averaging approximation. In contrast, the coupled Lorenz system necessitated a self-consistent approach instead of the averaging method. In this paper, we extend our previous results for the coupled Lorenz system and provide detailed calculations. We analytically and numerically show that, under strong coupling, WP and TL with an exponent 2 emerge not only in the globally coupled Lorenz system~\cite{Lee_etal_1998_PRL}, which is based on the original Lorenz system \cite{Lorenz_1963}, but also in globally coupled systems constructed from the generalized Lorenz system~\cite{Macek_Strumik_2010_PRE, Macek_Strumik_2014_PRL} and hyperchaotic Lorenz system~\cite{Wang_Wang_2008_PhysicaA}. The generalized Lorenz system, an extension of the original Lorenz system that incorporates the influence of a magnetic field \cite{Macek_Strumik_2010_PRE}, has been shown to exhibit intriguing intermittent dynamics \cite{Macek_Strumik_2014_PRL}. The hyperchaotic Lorenz system \cite{Wang_Wang_2008_PhysicaA}, also an extension of the original Lorenz system, displays more complex dynamics than the original Lorenz system \cite{Lorenz_1963} and is of particular interest for applications in secure communication and cryptography. Moreover, we derive a condition under which WP and TL with an exponent 2 hold in the original Lorenz system on networks. Using several representative network structures, such as a two-dimensional (2D) lattice, random networks, and scale-free networks, we numerically demonstrate that WP and TL with an exponent 2 emerge in strongly coupled network systems.

\section{Original Lorenz system}
\subsection{Model}
First, we present the results for the original Lorenz system~\cite{Lorenz_1963}:
\begin{subequations}
\label{eq:Lorenz}
\begin{align}
\dot{x} &= \sigma (y - x), \\
\dot{y} &= x(\rho - z) - y, \\
\dot{z} &= x y - b z.
\end{align}
\end{subequations}
Using this system, we construct the following coupled system~\cite{Lee_etal_1998_PRL}:
\begin{subequations}
\label{eq:coupled_Lorenz}
\begin{align}
\dot{x}_i &= \sigma (y_i- x_i) + D_x(X-x_i), \label{eq:coupled_Lorenz_x} \\
\dot{y}_i &= x_i(\rho_i-z_i) - y_i + D_y(Y-y_i), \label{eq:coupled_Lorenz_y} \\
\dot{z}_i &= x_i y_i - bz_i + D_z(Z-z_i),\label{eq:coupled_Lorenz_z}
\end{align}
\end{subequations}
where $D_x$, $D_y$, and $D_z$ are coupling coefficients, and $X = \langle x_i \rangle_i$, $Y = \langle y_i \rangle_i$, and $Z = \langle z_i \rangle_i$. The notation $\langle \cdot \rangle_i$ denotes the ensemble average, namely, $\langle v_i \rangle_i = (1/N)\sum_{i=1}^{N} v_i \ (v_i = x_i, y_i, z_i).$ The parameter $\rho_i$ is defined as
\begin{subequations}
\label{eq:hetero_para}
\begin{align}
    \rho_i &= \rho_0 + \mu_i, \label{eq:rho_i}\\
    \langle \mu_i \rangle_i &= 0. \label{eq:mu_i}
\end{align}
\end{subequations}
Equation \eqref{eq:hetero_para} is satisfied by setting $\rho_0 = \langle \rho_i \rangle_i$.

 In Sec.~II B, we analytically demonstrate that, under strong coupling, WP and TL (defined below) arise in the variable $z_i(t)$. We consider a time-series set $z_i(t)$ $(i=1,\ldots,N)$. For temporal TL, we compute the mean and variance for each oscillator $i$ as ${\rm E}[z_i(t)]_t=\langle z_i(t) \rangle_t$ and ${\rm V}[z_i(t)]_t=\langle (z_i(t) - {\rm E}[z_i(t)]_t)^2 \rangle_t$, respectively, where $\langle \cdot \rangle_t$ denotes the long-time average or the average over one cycle when $z_i(t)$ is periodic. A linear fitting to $N$ data points of $(\log {\rm E}[z_i(t)]_t, \log {\rm V}[z_i(t)]_t)$ yields the slope $\beta_{\rm t}$ and intercept $\log \alpha_{\rm t}$. For spatial TL, the mean and variance at time $t$ are given by ${\rm E}[z_i(t)]_i = \langle z_i(t) \rangle_i$ and ${\rm V}[z_i(t)]_i=\langle (z_i(t) - {\rm E}[z_i(t)]_i)^2 \rangle_i$, respectively, where $\langle \cdot \rangle_i$ denotes the average over oscillator $i$. A linear fitting to $M$ data points $(\log {\rm E}[z_i(t)]_i, \log {\rm V}[z_i(t)]_i)$, where $M$ is the number of sample times, yields the slope $\beta_{\rm s}$ and intercept $\log \alpha_{\rm s}$. In both cases, $R^2$ denotes the coefficient of determination for the linear fitting. When WP holds, TL with an exponent 2 naturally arises, as shown below. WP is defined by the following relation:
\begin{eqnarray}
    z_i(t) = C_i z_0(t). \label{eq:WP}
\end{eqnarray}
Under this condition, the temporal mean, temporal variance, ensemble mean, and ensemble variance of $z_i(t)$ are given by
\begin{eqnarray}
        {\rm E}[z_i(t)]_t &=& C_i\, {\rm E}[z_0(t)]_t, \label{eq:temporal_mean_z_i}\\
        {\rm V}[z_i(t)]_t &=& C_i^{\, 2}\, {\rm V}[z_0(t)]_t, \label{eq:temporal_var_z_i} \\
        {\rm E}[z_i(t)]_i &=& {\rm E}[C_i]_i\, z_0(t), \label{eq:ensemble_mean_z_i}\\
        {\rm V}[z_i(t)]_i &=& {\rm V}[C_i]_i\, \left[z_0(t)\right]^2. \label{eq:ensemble_var_z_i}
\end{eqnarray}
By eliminating $C_i$ from Eqs.~\eqref{eq:temporal_mean_z_i} and~\eqref{eq:temporal_var_z_i}, and $z_0(t)$ from Eqs.~\eqref{eq:ensemble_mean_z_i} and~\eqref{eq:ensemble_var_z_i}, we obtain the following relationships:
\begin{align}
\mathrm{V}[z_i(t)]_t &= \dfrac{\mathrm{V}[z_0(t)]_t}{\mathrm{E}[z_0(t)]_t^{\, 2}}\, \mathrm{E}[z_i(t)]_t^{\, 2}, \label{eq:temporalTL} \\
\mathrm{V}[z_i(t)]_i &= \dfrac{\mathrm{V}[C_i]_i}{\mathrm{E}[C_i]_i^{\, 2}}\, \mathrm{E}[z_i(t)]_i^{\, 2}. \label{eq:spatialTL}
\end{align}
These equations represent temporal and spatial TLs with an exponent 2, respectively. In practice, WP, as given by Eq.~\eqref{eq:WP}, may include a time delay and is more accurately described by
\begin{eqnarray}
    z_i(t) = C_i z_0(t-t_i), \label{eq:WP_modified}
\end{eqnarray}
where $t_i$ is a time delay that depends on the coupling strength and decreases as the coupling becomes stronger. Under the condition described by Eq.~\eqref{eq:WP_modified}, the relation~\eqref{eq:temporalTL} still holds, whereas the relation~\eqref{eq:spatialTL} does not. Therefore, the emergence of spatial TL requires stronger coupling than that for temporal TL~\cite{Mitsui_Kori_2025_PRL}.

\subsection{Analytical derivation}
To demonstrate that WP and TL with an exponent 2 hold for $z_i(t)$, we approximately solve the coupled Lorenz system described by Eq.~(\ref{eq:coupled_Lorenz}). As ansatzes, we consider the following expansion~\cite{Mitsui_Kori_2025_PRL}:
\begin{subequations}
  \label{eq:ansatz}
\begin{align}
 x_i(t) &= x_0(t- \eps_i \tau) + \eps_i p(t- \eps_i \tau) + O(\hat{\eps}^2), \label{x_ansatz}\\
 y_i(t) &= y_0(t- \eps_i \tau) + \eps_i q(t- \eps_i \tau) + O(\hat{\eps}^2), \label{y_ansatz}\\
 z_i(t) &= z_0(t- \eps_i \tau) + \eps_i r(t- \eps_i \tau) + O(\hat{\eps}^2), \label{z_ansatz}
\end{align}
\end{subequations}
where $\eps_i = \mu_i / D_x$ is a nondimensional small parameter; $\tau$ is a constant; $p(t)$, $q(t)$, and $r(t)$ are functions to be determined; $\hat{\eps}$ represents the typical magnitude of the small parameter $\eps_i$; and $x_0$, $y_0$, and $z_0$ constitute a reference oscillator satisfying
\begin{subequations}
\label{eq:refosci_Lorenz}
\begin{align}
\dot{x}_0 &= \sigma (y_0 - x_0), \label{eq:dot_x0}\\
\dot{y}_0 &= x_0(\rho_0 - z_0) - y_0, \label{eq:dot_y0}\\
\dot{z}_0 &= x_0 y_0 - b z_0. \label{eq:dot_z0}
\end{align}
\end{subequations}
Our objective is to show that $r(t) \propto z_0(t)$ holds in a good approximation for a sufficiently small $\hat \eps$ and sufficiently large $D_x$. If $r(t) \propto z_0(t)$ is approximately true, WP, i.e., $z_1(t) \propto \ldots  \propto z_N(t)$, also holds approximately.
When $|\sigma(y_0-x_0)| \gg |\mu_ip|, |\eps_i \sigma q|, |\mu_i\tau\sigma(y_0-x_0)|$, by substituting Eq.~(\ref{eq:ansatz}) into Eq.~(\ref{eq:coupled_Lorenz}) and extracting $O(\eps_i)$ terms from both sides, we obtain
\begin{subequations}
\label{eq:dot_pqrs}
\begin{align}
\dot{p} =& -(\sigma + D_x)p + \sigma q + D_x\tau\sigma(y_0-x_0), \label{eq:dot_p} \\
\dot{q} =& -(1 + D_y)q + (D_x-r)x_0 + (\rho_0-z_0)p \nonumber \\ 
&+ D_y\tau x_0(\rho_0-z_0)-D_y\tau y_0, \label{eq:dot_q}\\
\dot{r} =& -(b + D_z)r + py_0 + qx_0 + D_z\tau x_0 y_0 - D_z\tau b z_0.\label{eq:dot_r}
\end{align}
\end{subequations}
When $D_x$ is sufficiently large, $p$ is a fast variable, and we may adiabatically eliminate $p$ from the system in a good approximation. Solving $\dot p=0$, we find
\begin{align}
    p = \dfrac{\sigma q + D_x\tau\sigma(y_0 - x_0)}{\sigma + D_x}. \label{eq:p}
\end{align}
Substituting Eq.~(\ref{eq:p}) into Eqs.~\eqref{eq:dot_q} and \eqref{eq:dot_r}, assuming that $q$ and $r$ are $O(D_x)$, and then eliminating terms smaller than $O(D_x)$, Eqs.~\eqref{eq:dot_q} and \eqref{eq:dot_r} can be approximated to
\begin{subequations}
\label{eq:dot_qr2}
\begin{align}
    \dot{q} =& -(1+D_y) q + (D_x - r)x_0, \label{eq:dot_q2} \\
    \dot{r} =& -(b + D_z) r + q x_0. \label{eq:dot_r2}
\end{align}
\end{subequations}
Assuming that $r(t)$ is $O(D_x)$, Eq.~\eqref{eq:dot_q2} implies that $q(t)$ is also $O(D_x)$. Then, Eq.~\eqref{eq:dot_r2} confirms that $r(t)$ is indeed $O(D_x)$. We now consider the case where there are no couplings in the variables $y_i$ and $z_i$ (i.e., $D_y = D_z = 0$), or when the couplings are weak ($D_y \ll 1$ and $D_z \ll b$). In these cases, the equations simplify to
\begin{subequations}
\label{eq:dot_qr3}
\begin{align}
    \dot{q} =& - q + (D_x - r)x_0, \label{eq:dot_q3} \\
    \dot{r} =& - b r + q x_0. \label{eq:dot_r3}
\end{align}
\end{subequations}
With Eqs.~(\ref{eq:dot_y0}), (\ref{eq:dot_z0}), (\ref{eq:dot_q3}), and (\ref{eq:dot_r3}), we can confirm that the following proportional relationships hold:
\begin{align}
q(t) = & \dfrac{D_x}{\rho_0} y_0(t), \label{eq:q_propto_y0} \\
r(t) = & \dfrac{D_x}{\rho_0} z_0(t). \label{eq:r_propto_z0}
\end{align}
Indeed, by substituting Eqs.~\eqref{eq:q_propto_y0} and \eqref{eq:r_propto_z0} into Eq.~\eqref{eq:dot_qr3} and rearranging the terms with the aid of Eq.~\eqref{eq:refosci_Lorenz}, we can verify that Eqs.~\eqref{eq:q_propto_y0} and \eqref{eq:r_propto_z0} are solutions of Eq.~\eqref{eq:dot_qr3}. Thus, $r(t) \propto z_0(t)$ is approximately true, which implies that WP holds for $z_i(t)$. Consequently, the theoretical predictions for the intercepts and exponents of temporal and spatial TLs are given by
\begin{subequations}
\label{eq:TL_theory}
\begin{align}
\log \alpha_{\mathrm{t}} &= \log \dfrac{\mathrm{V}[z_0(t)]_t}{\mathrm{E}[z_0(t)]_t^{\, 2}}, \\
\beta_{\mathrm{t}} &= 2, \\
\log \alpha_{\mathrm{s}} &= \log \dfrac{\mathrm{V}[\mu_i]_i}{\rho_0^{\, 2}}, \\
\beta_{\mathrm{s}} &= 2.
\end{align}
\end{subequations}
Since TL is typically discussed on a logarithmic scale, it is defined only for variables that take positive values. Accordingly, in this study, we mainly focus on the variable $z_i(t)$, which is always positive. In contrast, from Eq.~\eqref{eq:q_propto_y0}, we find that $q(t) \propto y_0(t)$, indicating that WP also holds for the variable $y_i(t)$. Therefore, it is possible to derive the quadratic relationships between the mean and variance of $y_i(t)$ as follows:
\begin{subequations}
\label{eq:TL_y}
\begin{align}
\mathrm{V}[y_i(t)]_t &=  \dfrac{\mathrm{V}[y_0(t)]_t}{\mathrm{E}[y_0(t)]_t^{\, 2}}\mathrm{E}[y_i(t)]_t^{\, 2}, \label{eq:temporalTL_y}\\
\mathrm{V}[y_i(t)]_i &= \dfrac{\mathrm{V}[\mu_i]_i}{\rho_0^{\, 2}}\mathrm{E}[y_i(t)]_i^{\, 2}. \label{eq:spatialTL_y}
\end{align}
\end{subequations}
However, in numerical simulations, the temporal mean of $y_0(t)$ approaches $0$, causing the denominator in Eq.~\eqref{eq:temporalTL_y} to vanish. As a result, defining temporal TL for the variable $y_i(t)$ is practically difficult.
\subsection{Numerical simulations}
In this section, we validate the analytical derivation through numerical simulations. Before proceeding, we introduce two quantities, $\chi$ and $\zeta_i(t)$, to quantify the degree of synchronization between time series and to assess the presence of WP, respectively:
\begin{eqnarray}
\chi &=& \dfrac{\mathrm{CV}[Z(t)]}{\underset{i}{\max}\{\mathrm{CV}[z_i(t)]\}},\\
\zeta_i(t) &=& \dfrac{z_i(t)}{z_j(t)},
\end{eqnarray}
where $\mathrm{CV}$ denotes the coefficient of variation. That is, $\chi$ is the CV of the mean field of $z_i(t)$ normalized by the maximum CV among the individual $z_i(t)$. With this definition, $0 \leq \chi \leq 1$, where $\chi \simeq 0$ indicates an unsynchronized state and $\chi \simeq 1$ signifies a well-synchronized state. The quantity $\zeta_i(t)$ represents the ratio between time series; it becomes approximately constant when WP holds. While the choice of $j$ is arbitrary, we select $j = 50$, assuming that the parameters $\mu_i$ satisfy $\mu_1 < \mu_2 < \cdots < \mu_{100}$. By arranging them in this manner, we can expect $z_{50}(t) \simeq z_0(t)$. Unless otherwise noted, $\mu_i$ is selected from a uniform distribution between $-1.5$ and $1.5$.

\begin{figure*}[htb]
\includegraphics[width=160mm]{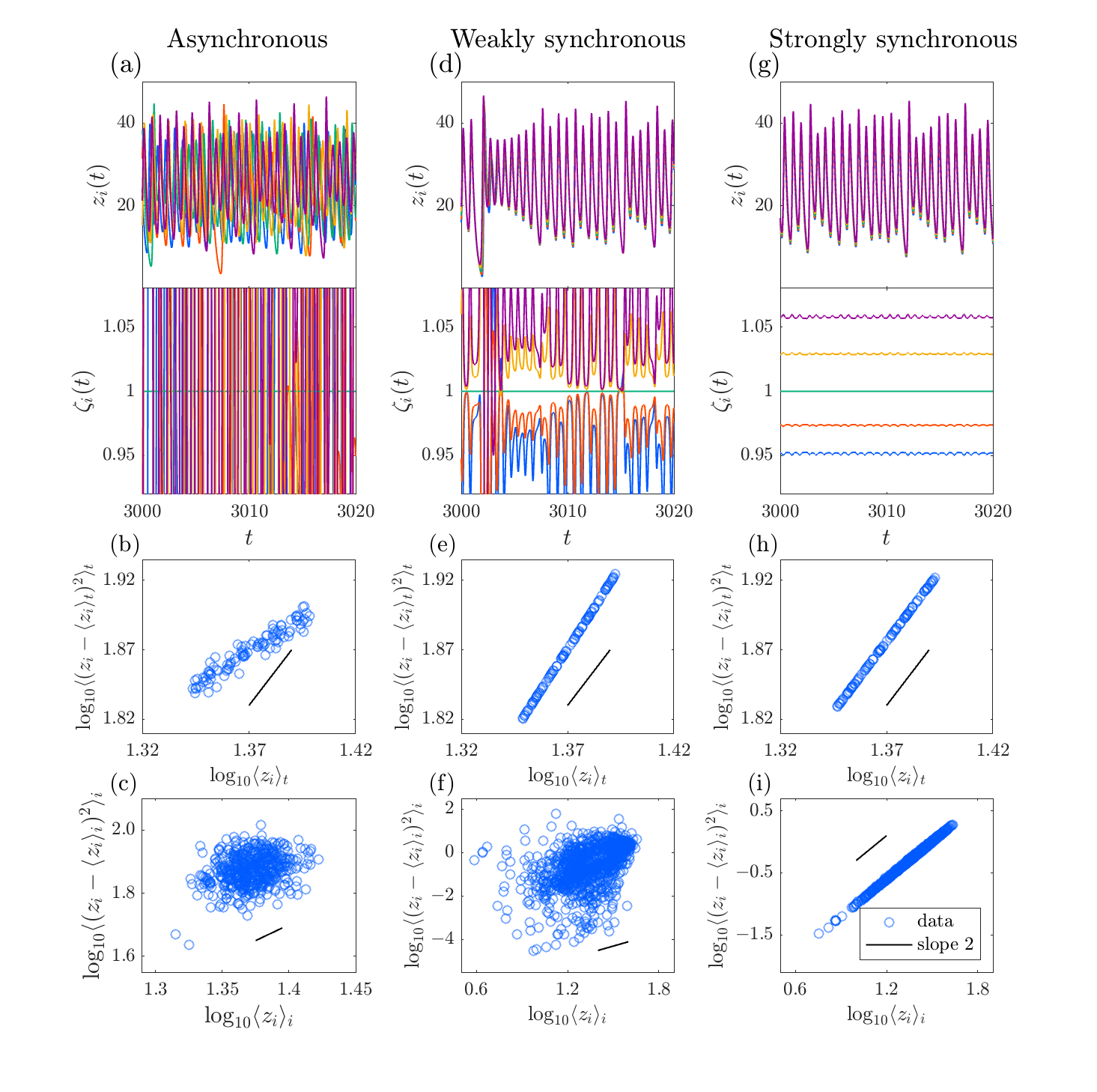}
\caption{\label{fig:Lorenz_async_sync} Examples of WP and TL in the globally coupled Lorenz system. $N=100, D_y = D_z = 0, \sigma = 10$, $b = 8/3$, and $\rho_i$ is randomly selected from a uniform distribution between $26.5$ and $29.5$. In panels (a), (d), and (g), data for $i=1, 25, 50, 75,$ and $100$ are shown. (a) $z_i(t)$ (top) and $\zeta_i(t)$ (bottom) for $D_x=0$. (b) Temporal TL for $D_x=0$. (c) Spatial TL for $D_x=0$. (d) $z_i(t)$ (top) and $\zeta_i(t)$ (bottom) for $D_x=20$. (e) Temporal TL for $D_x=20$. (f) Spatial TL for $D_x=20$. (g) $z_i(t)$ (top) and $\zeta_i(t)$ (bottom) for $D_x=1000$. (h) Temporal TL for $D_x=1000$. (i) Spatial TL for $D_x=1000$.}
\end{figure*}

First, for a sufficiently large $D_x$, we confirm that WP and TL with an exponent 2 are observed in the globally coupled Lorenz system. Figure~\ref{fig:Lorenz_async_sync} shows that without coupling, the time series are asynchronous, WP does not hold, and neither temporal TL nor spatial TL with an exponent 2 is observed. At an intermediate coupling strength, the system exhibits a weakly synchronous state in which only temporal TL appears. This indicates that spatial TL requires stronger coupling to emerge than temporal TL. With sufficiently strong coupling, the time series become strongly synchronous, WP holds, and both temporal and spatial TLs with an exponent 2 are clearly observed.

\begin{figure*}[htbp]
\includegraphics[width=180mm]{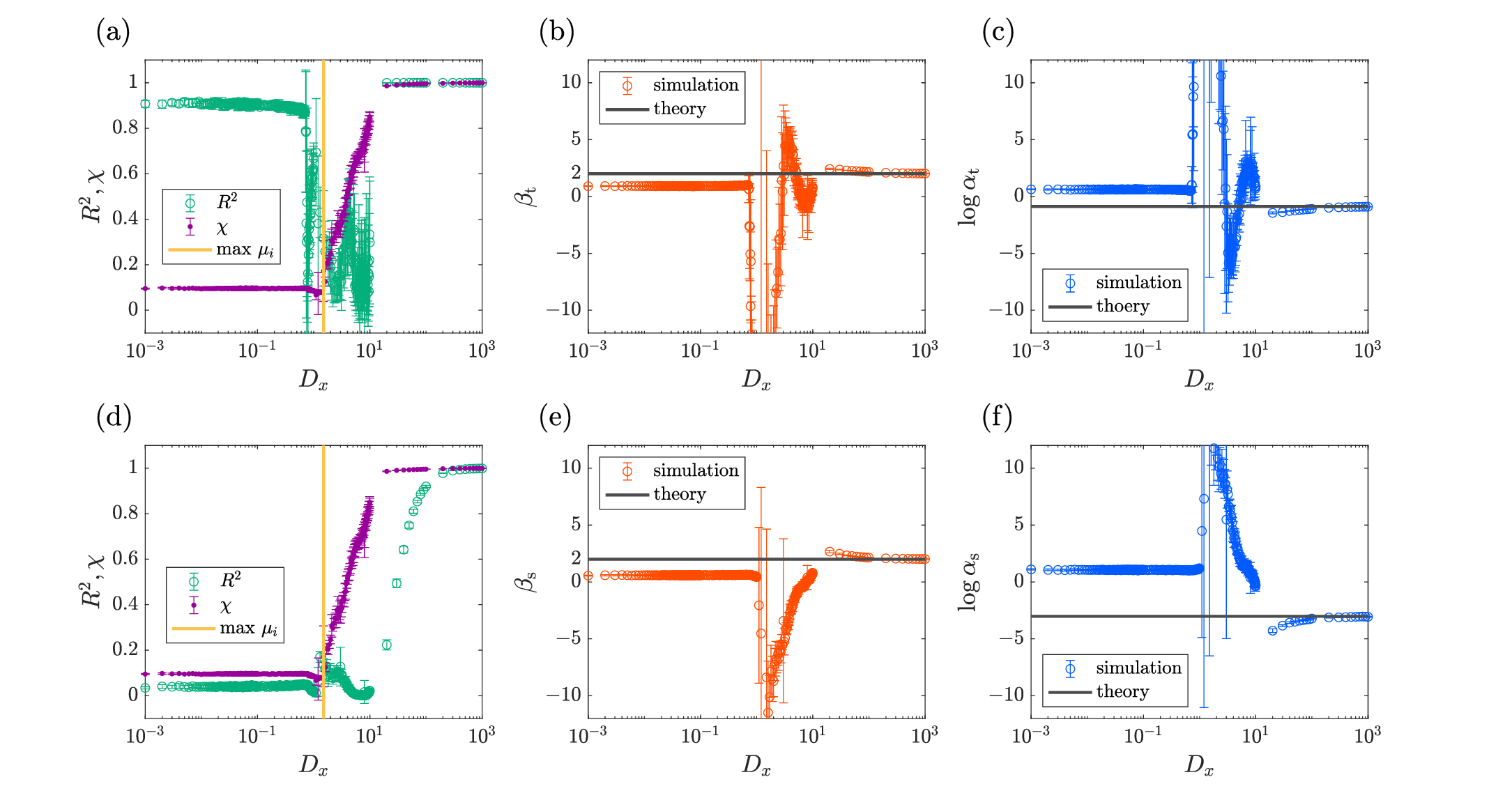}
\caption{\label{fig:Lorenz_TL} Dependence of TL parameters and synchronization degree on $D_x$ in the globally coupled Lorenz system. $D_y = D_z= 0, N=100, \sigma = 10$, $b = 8/3$, and $\rho_i$ is randomly selected from a uniform distribution between $26.5$ and $29.5$. In panels (a) and (d), the same $\chi$ (purple dot) and $\mbox{max}\{\mu_i\}$ (orange solid line) are plotted. (a) $R^2$ for
temporal TL. (b) Exponents of temporal TL. (c) Intercepts of temporal TL. (d) $R^2$ for spatial TL. (e) Exponents of spatial TL. (f) Intercepts of spatial TL.}
\end{figure*}

Next, we examine the dependence of the TL parameters on the coupling strength. Based on the analytical derivation, the numerical simulation results are expected to agree well with the theoretical predictions as $D_x$ increases. The results are shown in Fig.~\ref{fig:Lorenz_TL}. Simulations are performed up to $t=3500$, and TL is computed using the time series from $t=3000$ to $t=3500$. The error bars represent the standard deviation of ten calculations with different initial conditions and realizations of $\rho_i$. The initial conditions for the numerical simulations are randomly chosen from a uniform distribution between 0 and 1. Depending on the coupling strength, oscillation quenching may occur, rendering it impossible to define $\chi$. Thus, we judge that oscillation quenching occurs when $(1/N) \sum_{i=1}^{N}\langle\left(z_i-\langle z_i\rangle_t\right)^2\rangle_t$ falls below a certain threshold value; such cases are excluded. We can confirm that qualitatively the same results are obtained for several different threshold values. This procedure is applied to all systems described later.
Figure~\ref{fig:Lorenz_TL} illustrates that, when the coupling strength $D_x$ becomes comparable to $\max \{\mu_i\}$, $\chi$ starts to increase, indicating the onset of synchronization. With a further increase in the coupling strength, $\chi \simeq 1$ and $R^2 \simeq 1$, and the theoretical predictions and numerical simulation results for both $\log \alpha_{\rm t,s}$ and $\beta_{\rm t,s}$ show excellent agreement.

\section{Extensions}
In this section, we show that temporal and spatial TLs with an exponent 2 hold in network systems and high-dimensional Lorenz systems. 
\subsection{Network systems}
\begin{figure*}[t]
\includegraphics[width=180mm]{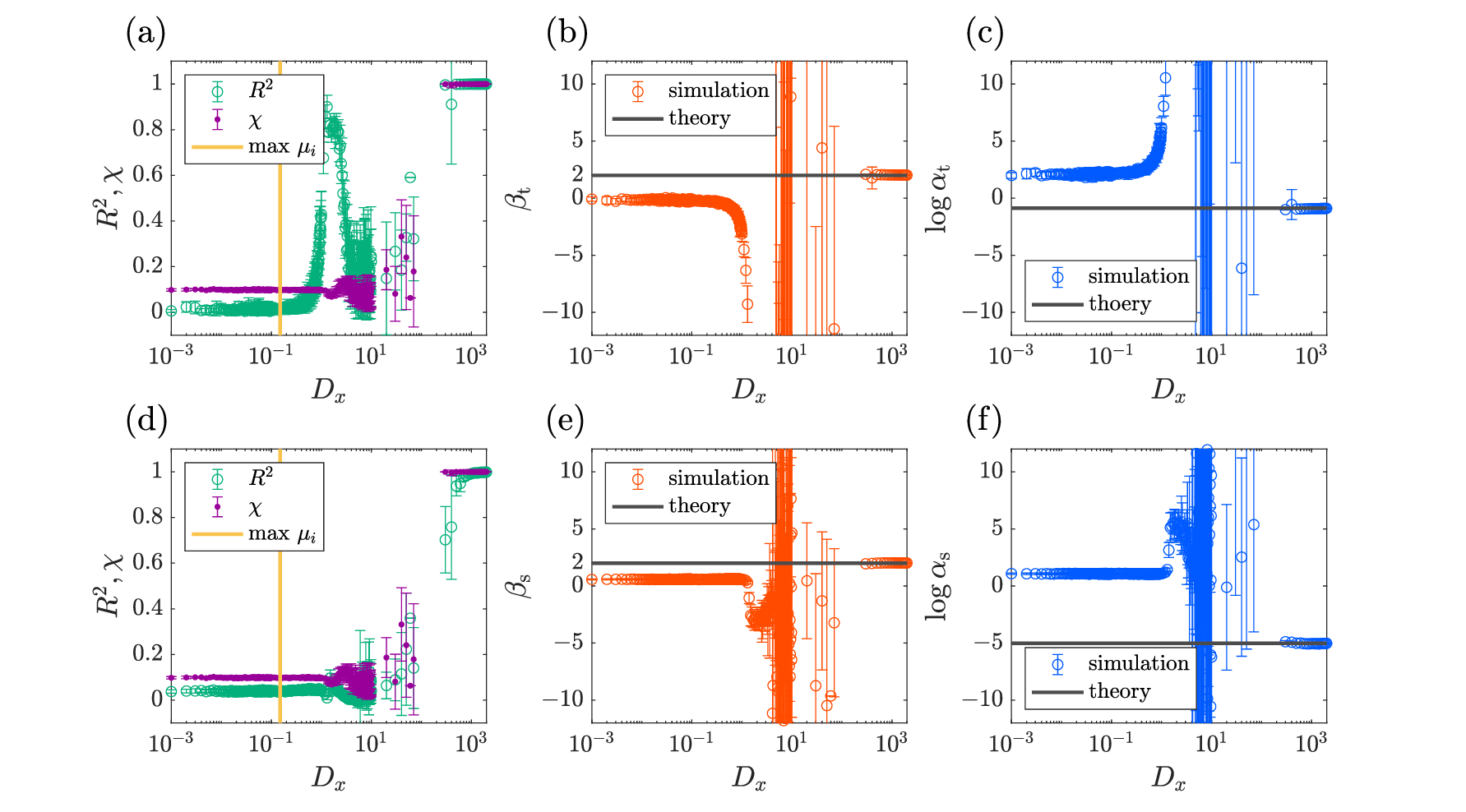}
\caption{\label{fig:Lorenz_TL_2dim_lattice} Dependence of TL parameters and synchronization degree on $D_x$ in the coupled Lorenz system on a 2D lattice. The side length of the 2D lattice is $10$. $N=100, \sigma = 10$, $b = 8/3$, and $\rho_i$ is randomly selected from a uniform distribution between $27.85$ and $28.15$ (namely, $\mu_i$ is selected from a uniform distribution between $-0.15$ and $0.15$). In panels (a) and (d), the same $\chi$ (purple dot) and $\mbox{max}\{\mu_i\}$ (orange solid line) are plotted. (a) $R^2$ for
temporal TL. (b) Exponents of temporal TL. (c) Intercepts of temporal TL. (d) $R^2$ for spatial
TL. (e) Exponents of spatial TL. (f) Intercepts of spatial TL.}
\end{figure*}

\begin{figure*}[t]
\includegraphics[width=180mm]{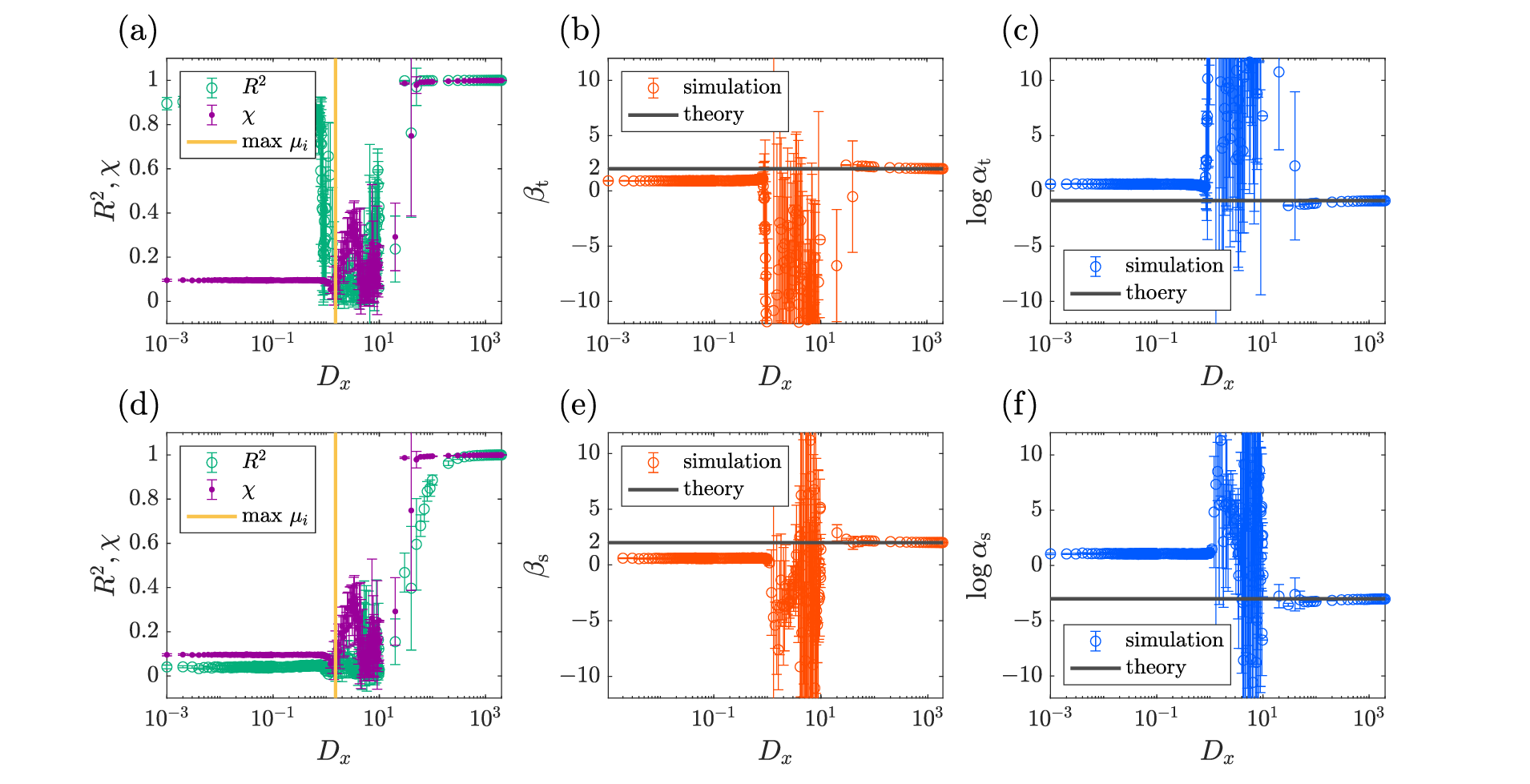}
\caption{\label{fig:Lorenz_TL_random_network} Dependence of TL parameters and synchronization degree on $D_x$ in the coupled Lorenz system on random networks. The probability of placing an edge is $p=0.05$. $N=100, \sigma = 10$, $b = 8/3$, and $\rho_i$ is randomly selected from a uniform distribution between $26.5$ and $29.5$. In panels (a) and (d), the same $\chi$ (purple dot) and $\mbox{max}\{\mu_i\}$ (orange solid line) are plotted. (a) $R^2$ for
temporal TL. (b) Exponents of temporal TL. (c) Intercepts of temporal TL. (d) $R^2$ for spatial
TL. (e) Exponents of spatial TL. (f) Intercepts of spatial TL.}
\end{figure*}

\begin{figure*}[t]
\includegraphics[width=180mm]{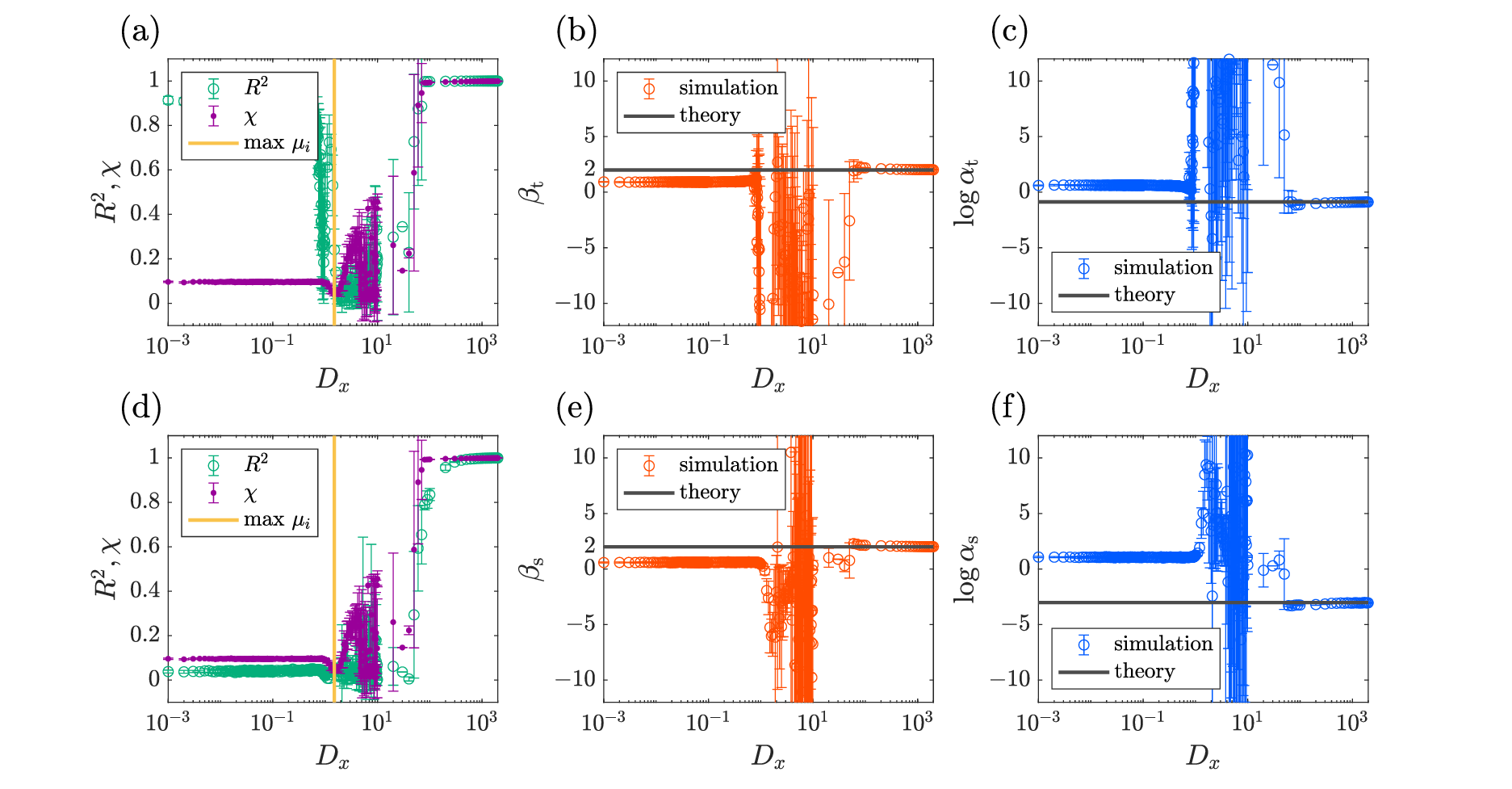}
\caption{\label{fig:Lorenz_TL_scalefree_network} Dependence of TL parameters and synchronization degree on $D_x$ in the coupled Lorenz system on scale-free networks. The scale-free networks are generated using the Barabási–Albert model with an initial network size $m_0 =3$, the number of newly attached edges per added node $m=2$, and a final network size $N=100$. $\sigma = 10$, $b = 8/3$, and $\rho_i$ is randomly selected from a uniform distribution between $26.5$ and $29.5$. In panels (a) and (d), the same $\chi$ (purple dot) and $\mbox{max}\{\mu_i\}$ (orange solid line) are plotted. (a) $R^2$ for 
temporal TL. (b) Exponents of temporal TL. (c) Intercepts of temporal TL. (d) $R^2$ for spatial
TL. (e) Exponents of spatial TL. (f) Intercepts of spatial TL.}
\end{figure*}

Thus far, we have presented the results for the globally coupled system. In the strong-coupling regime, however, we also found that network systems exhibit TL with an exponent 2. We introduce the Lorenz system on networks as follows:
\begin{subequations}
\label{eq:network_Lorenz}
\begin{align}
\dot{x}_i &= \sigma (y_i- x_i) + \dfrac{D_x}{\kappa_i}\sum_{j=1}^{N}A_{ij}(x_j-x_i), \label{eq:network_Lorenz_x} \\
\dot{y}_i &= x_i(\rho_i-z_i) - y_i, \label{eq:network_Lorenz_y} \\
\dot{z}_i &= x_i y_i - bz_i,\label{eq:network_Lorenz_z}
\end{align}
\end{subequations}
where $D_x$ is the coupling strength, $\kappa_i$ is the degree of node $i$, and $A_{ij}$ is the adjacency matrix representing network structures. The parameter $\rho_i$ is defined as Eq.~\eqref{eq:hetero_para}. In this case, we obtain the following equation by the perturbative calculations:
\begin{align}
    \eps_i \dot{p} + O(\hat{\eps}^2) =&
    \eps_i\sigma(q - p) +  \left(\dfrac{1}{\kappa_i}\sum_{j=1}^{N}A_{ij}\eps_j - \eps_i\right) \nonumber \\
    &\times D_x ( p + \eps_i\tau\dot{p} - \tau\dot{x}_0 - \eps_i\tau^2\ddot{x}_0 ) + O(\hat{\eps}^2).
\end{align}
If $(1/\kappa_i)\sum_{j=1}^{N}A_{ij}\eps_j$ is small and the following condition
\begin{eqnarray}
    \dfrac{1}{\kappa_i} \sum_{j=1}^{N}A_{ij}\eps_j - \eps_i \simeq -\eps_i, \label{eq:condition_network}
\end{eqnarray}
is satisfied, we obtain
\begin{align}
    \eps_i \dot{p} + O(\hat{\eps}^2) =&
    \eps_i\sigma(q - p)  \nonumber - \eps_i  D_x \\ &\times( p + \eps_i\tau\dot{p} - \tau\dot{x}_0 - \eps_i\tau^2\ddot{x}_0 ) + O(\hat{\eps}^2),\\
     =& \eps_i\left[ \sigma q - \sigma p - D_xp + D_x\tau \sigma(y_0 - x_0) \right] \nonumber \\
     & + O(\hat{\eps}^2).
\end{align}
Extracting $O(\eps_i)$ terms from both sides of the equation, we obtain
\begin{align}
\dot{p} = -(\sigma + D_x)p + \sigma q + D_x\tau\sigma(y_0-x_0).
\end{align}
This evolution equation is identical to that of the globally coupled case; thus, the same proportional relations follow. We expect the condition \eqref{eq:condition_network} is satisfied when the oscillators are randomly distributed within the network. We confirm this expectation using a 2D lattice system, random networks, and scale-free networks. As shown in Figs.~\ref{fig:Lorenz_TL_2dim_lattice}, \ref{fig:Lorenz_TL_random_network}, and \ref{fig:Lorenz_TL_scalefree_network}, the exponents of TL approach 2 in the large $D_x$ regime, and the intercepts also show good agreement with the theoretical predictions.

Next, we consider a case where the condition \eqref{eq:condition_network} is not satisfied. For example, in a one-dimensional (1D) lattice system where the oscillators are arranged such that the values of the heterogeneity parameter increase in ascending order, the condition \eqref{eq:condition_network} is not satisfied, and instead the following relation is established:
\begin{eqnarray}
    \dfrac{1}{\kappa_i}\sum_{j=1}^{N}A_{ij}\eps_j - \eps_i \simeq 0. \label{eq:condition_network_breakdown}
\end{eqnarray}
In this case, the same analytical calculation procedure as in the globally coupled system cannot be applied. The numerical simulation results for this system are presented in Appendix~B. As shown in Appendix B, oscillation quenching occurs, making it impossible to evaluate WP and TL. Analyzing the mechanism behind this phenomenon would require examining the effects of $O(\eps_i^{\, 2})$ terms, which is beyond the scope of the present study.

\subsection{Generalized and hyperchaotic
Lorenz systems}
To assess the generality of the mechanism underlying the emergence of WP and TL with an exponent 2 identified in the original Lorenz system, we extend our focus to the generalized and hyperchaotic Lorenz systems and investigate whether the same phenomena arise in these models. The generalized Lorenz system~\cite{Macek_Strumik_2010_PRE, Macek_Strumik_2014_PRL} is described as follows:
\begin{subequations}
\label{eq:generalized_Lorenz}
\begin{align}
\dot{x} &= \sigma (y - x) - \omega_0 w,\\
\dot{y} &= x(\rho - z) - y, \\
\dot{z} &= x y - b z,\\
\dot{w} &= \omega_0 x - \sigma_{\rm m} w,
\end{align}
\end{subequations}
where $w(t)$, $\omega_0$, and $\sigma_{\rm m}$ represent a variable and parameters associated with the magnetic field. The hyperchaotic Lorenz system~\cite{Wang_Wang_2008_PhysicaA} is described as follows:
\begin{subequations}
\label{eq:hyperchaotic_Lorenz}
\begin{align}
\dot{x} &= \sigma (y - x) + w, \\
\dot{y} &= x(\rho - z) - y, \\
\dot{z} &= x y - b z,\\
\dot{w} &= -y z + \gamma w,
\end{align}
\end{subequations}
where $w(t)$ is a nonlinear controller and $\gamma$ is a newly introduced parameter. To treat these systems within a unified formulation, we introduce the following system:
\begin{subequations}
\label{eq:extended_Lorenz}
\begin{align}
\dot{x} &= \sigma (y - x) + aw, \\
\dot{y} &= x(\rho - z) - y, \\
\dot{z} &= x y - b z,\\
\dot{w} &= cx - kyz - lw,
\end{align}
\end{subequations}
where $x(t)$, $y(t)$, and $z(t)$ are the original variables of the Lorenz system and $\sigma$, $\rho$, and $b$ are its original parameters. The variable $w(t)$ and parameters $a$, $c$, $k$, and $l$ are newly introduced to extend the original Lorenz system. By appropriately choosing the parameters $a$, $c$, $k$, and $l$, we can recover the original/generalized/hyperchaotic Lorenz system; when $a = c = k = l = 0$, the system reduces to the original Lorenz system~\eqref{eq:Lorenz}; when $a = -\omega_0$, $c = \omega_0$, $k = 0$, and $l = \sigma_{\rm m}$, it corresponds to the generalized Lorenz system~\eqref{eq:generalized_Lorenz}; and when $a = 1$, $c = 0$, $k = 1$, and $l = -\gamma$, it corresponds to the hyperchaotic Lorenz system~\eqref{eq:hyperchaotic_Lorenz}. We refer to the system described by Eq.~(\ref{eq:extended_Lorenz}) as the extended Lorenz system. As will be described below, WP and associated TL with an exponent 2 can also be derived analytically for the coupled generalized and hyperchaotic Lorenz systems. Since the calculation procedures for these two systems are essentially identical, presenting them separately would be redundant. To avoid redundant derivations and simplify the presentation, we introduce the extended Lorenz system. Following the formulation of the coupled Lorenz system in the previous study~\cite{Lee_etal_1998_PRL}, we construct the following coupled system for the extended Lorenz system:
\begin{subequations}
\label{eq:coupled_extended_Lorenz}
\begin{align}
\dot{x}_i &= \sigma (y_i- x_i) + aw_i + D_x(X-x_i), \label{eq:coupled_extended_Lorenz_x} \\
\dot{y}_i &= x_i(\rho_i-z_i) - y_i + D_y(Y-y_i), \label{eq:coupled_extended_Lorenz_y} \\
\dot{z}_i &= x_i y_i - bz_i + D_z(Z-z_i),\label{eq:coupled_extended_Lorenz_z}\\
\dot{w}_i &= cx_i - ky_iz_i - lw_i + D_w(W-w_i), \label{eq:coupled_extended_Lorenz_w}
\end{align}
\end{subequations}
where $D_w$ is a coupling coefficient and $W = \langle w_i \rangle_i$. The parameter $\rho_i$ is defined as Eq.~\eqref{eq:hetero_para}. As in the original Lorenz system, we can analytically show that WP and TL with an exponent 2 also hold for the variable $z_i$ in the extended Lorenz system (see Appendix~A). As a result, the same proportional relationships as those in the original Lorenz system are also derived for the generalized and hyperchaotic Lorenz systems. In addition, depending on parameter values, we find that WP and TL with an exponent 2 hold for the variable $w_i(t)$ (see Appendix~A). However, since $w_i(t)$ takes both positive and negative values, TL in $w_i(t)$ is not a primary focus of the present study.

\begin{figure*}[t]
\includegraphics[width=180mm]{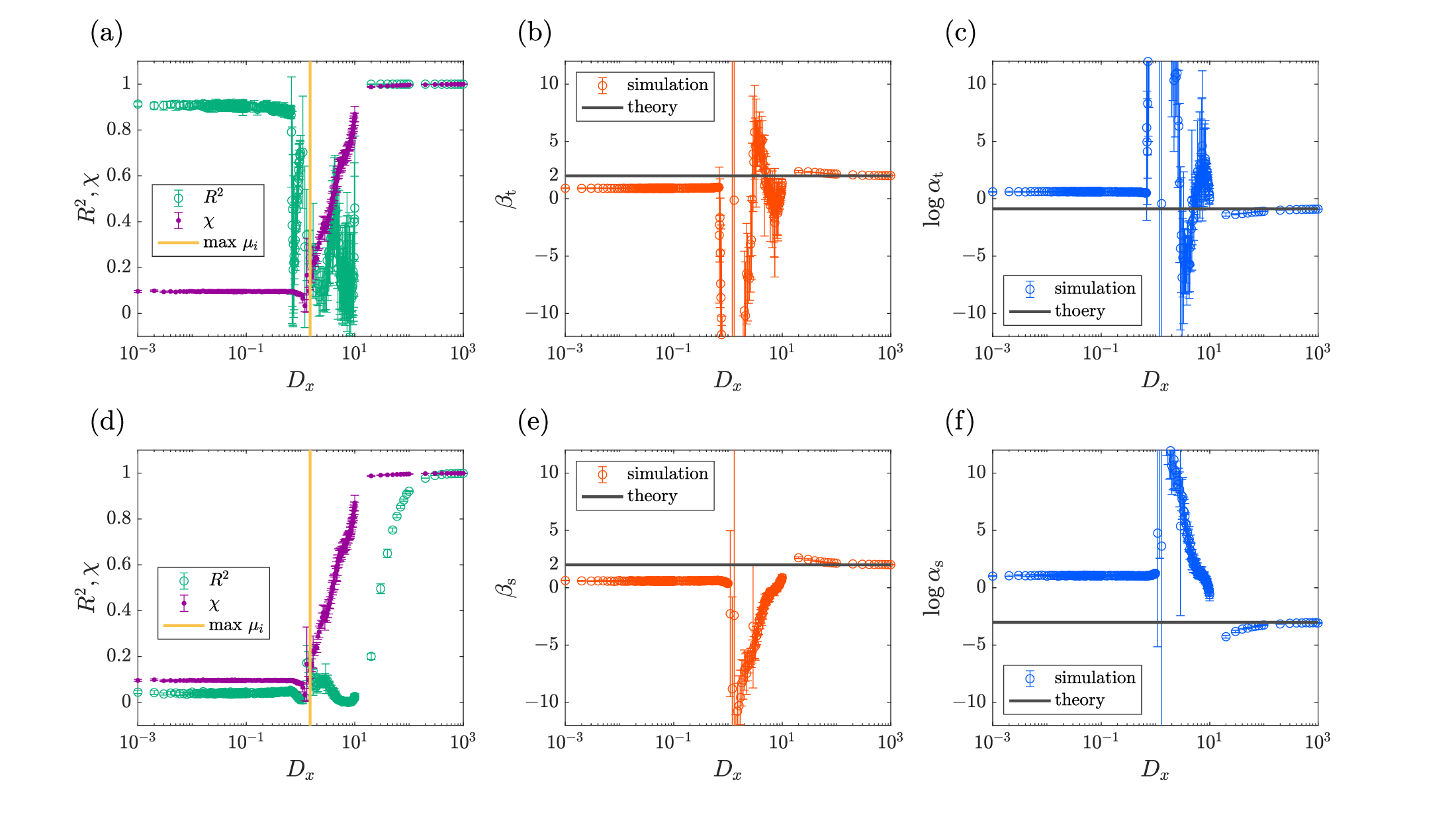}
\caption{\label{fig:generalized_Lorenz_TL} Dependence of TL parameters and synchronization degree on $D_x$ in the globally coupled generalized Lorenz system. $a = -1,  c = 1, k = 0, l = 20$, and $D_y = D_z = D_w = 0$. $N=100, \sigma = 10$, $b = 8/3$, and $\rho_i$ is randomly selected from a uniform distribution between $26.5$ and $29.5$. In panels (a) and (d), the same $\chi$ (purple dot) and $\mbox{max}\{\mu_i\}$ (orange solid line) are plotted. (a) $R^2$ for
temporal TL. (b) Exponents of temporal TL. (c) Intercepts of temporal TL. (d) $R^2$ for spatial
TL. (e) Exponents of spatial TL. (f) Intercepts of spatial TL.}
\end{figure*}

\begin{figure*}[t]
\includegraphics[width=180mm]{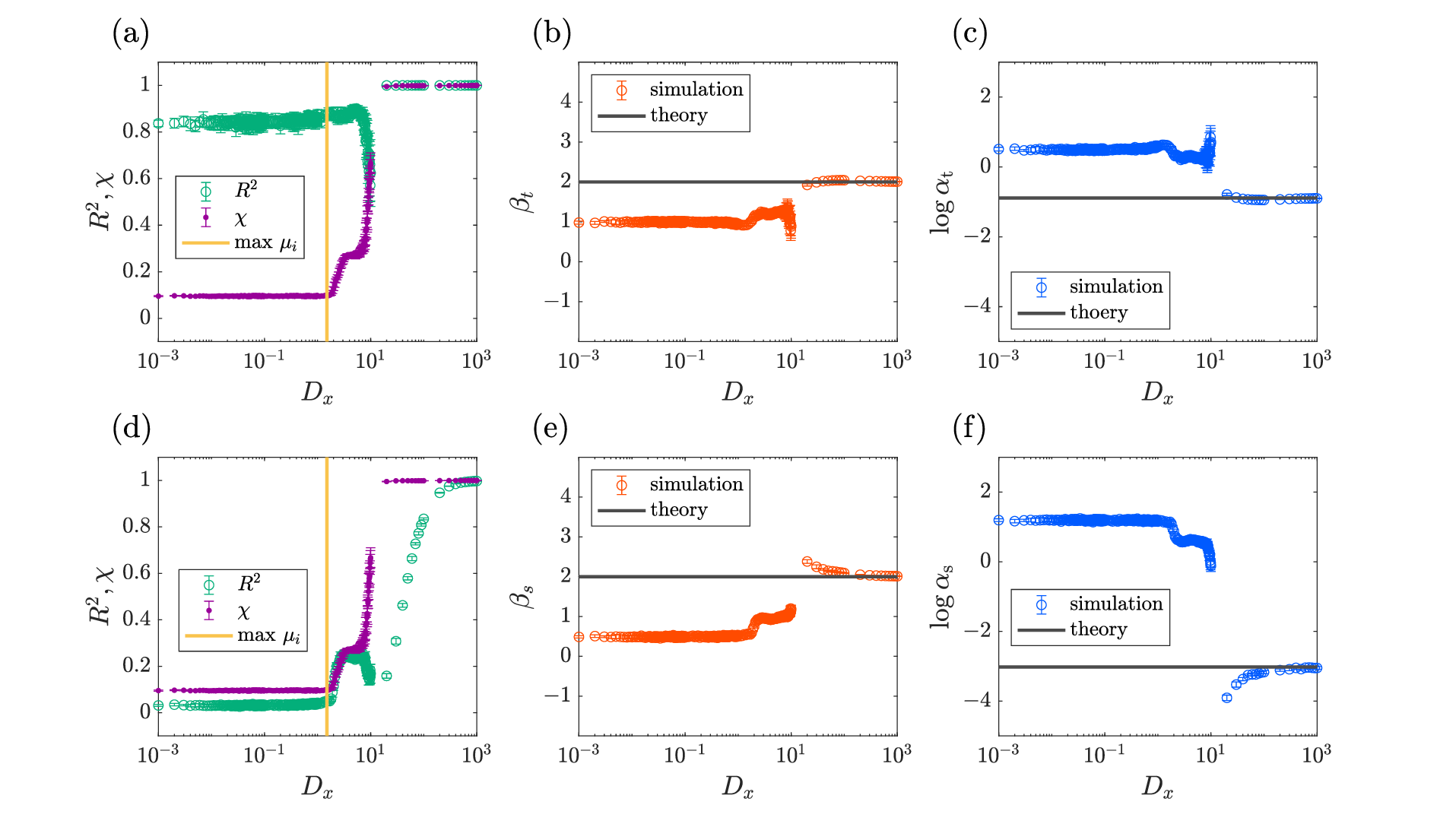}
\caption{\label{fig:hyperchaotic_Lorenz_TL} Dependence of TL parameters and synchronization degree on $D_x$ in the globally coupled hyperchaotic Lorenz system. $a = 1, c = 0, k = 1, l = 1$, and $D_y = D_z = D_w = 0$. $N=100, \sigma = 10$, $b = 8/3$, and $\rho_i$ is randomly selected from a uniform distribution between $26.5$ and $29.5$. In panels (a) and (d), the same $\chi$ (purple dot) and $\mbox{max}\{\mu_i\}$ (orange solid line) are plotted. (a) $R^2$ for
temporal TL. (b) Exponents of temporal TL. (c) Intercepts of temporal TL. (d) $R^2$ for spatial
TL. (e) Exponents of spatial TL. (f) Intercepts of spatial TL.}
\end{figure*}

As shown in Figs.~\ref{fig:generalized_Lorenz_TL} and \ref{fig:hyperchaotic_Lorenz_TL}, results similar to those obtained for the coupled Lorenz system are also observed in the coupled generalized Lorenz system and coupled hyperchaotic Lorenz system. In both cases, the exponent of TL approaches 2 in the large $D_x$ regime and the intercept shows good agreement with the theoretical prediction.

\begin{figure*}[htb]
\includegraphics[width=160mm]{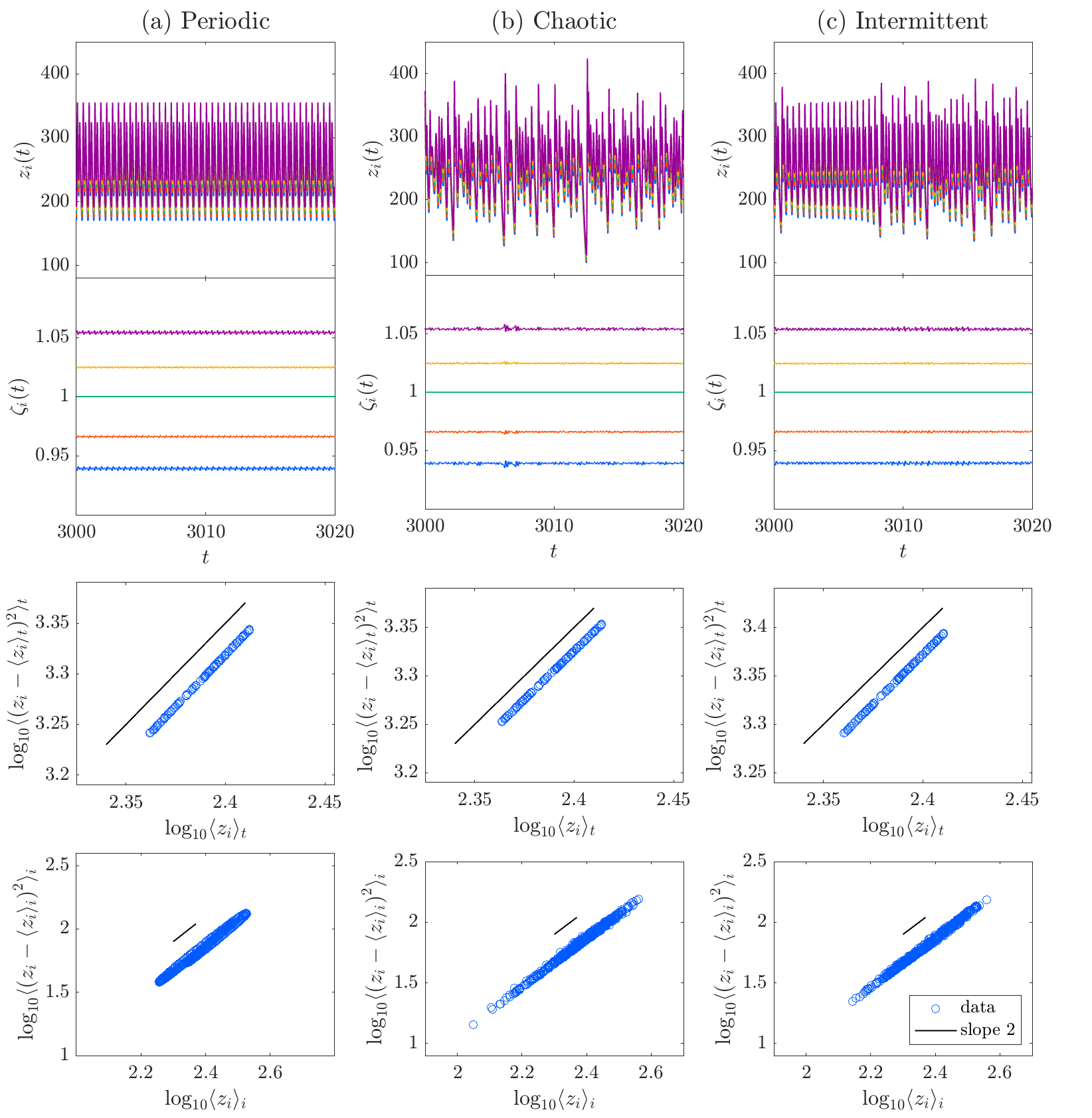}
\caption{\label{fig:generalized_Lorenz_WP_TL_example} Examples of WP and TL in the globally coupled generalized Lorenz system. $a = -\omega_0,  c = \omega_0, k = 0, l = 1$, $D_x = 1000$, and $D_y = D_z = D_w = 0$. $N=100, \sigma = 10$, $b = 8/3$, and $\rho_i$ is randomly selected from a uniform distribution between $241$ and $271$ (namely, $\mu_i$ is selected from a uniform distribution between $-15$ and $15$). For $z_i(t)$ and $\zeta_i(t)$, data
for $i = 1, 25, 50, 75$, and $100$ are shown. (a) Periodic: $\omega_0 = 1$. (b) Chaotic: $\omega_0 = 10$.  (c) Intermittent: $\omega_0 = 3.74$.}
\end{figure*}

Thus far, we have confirmed good agreement between the analytical predictions and numerical simulations for several models. In all cases, the system parameters were chosen such that the systems exhibited chaotic behavior. The results of the analytical calculations, however, do not depend on whether the system dynamics are chaotic or periodic. Therefore, using the coupled generalized Lorenz system, we verify that WP and TL with an exponent 2 also emerge when the system exhibits periodic behavior, in addition to chaotic dynamics. As the generalized Lorenz system is known to exhibit intermittent chaos for certain parameter values~\cite{Macek_Strumik_2014_PRL}, we also include the results obtained under parameter settings that lead to intermittent chaos. As shown in Fig.~\ref{fig:generalized_Lorenz_WP_TL_example}, WP and TL with an exponent 2 are clearly observed when $D_x$ is sufficiently large, regardless of whether the dynamics are periodic, chaotic, or intermittent.

\begin{figure*}[htbp]
\includegraphics[width=180mm]{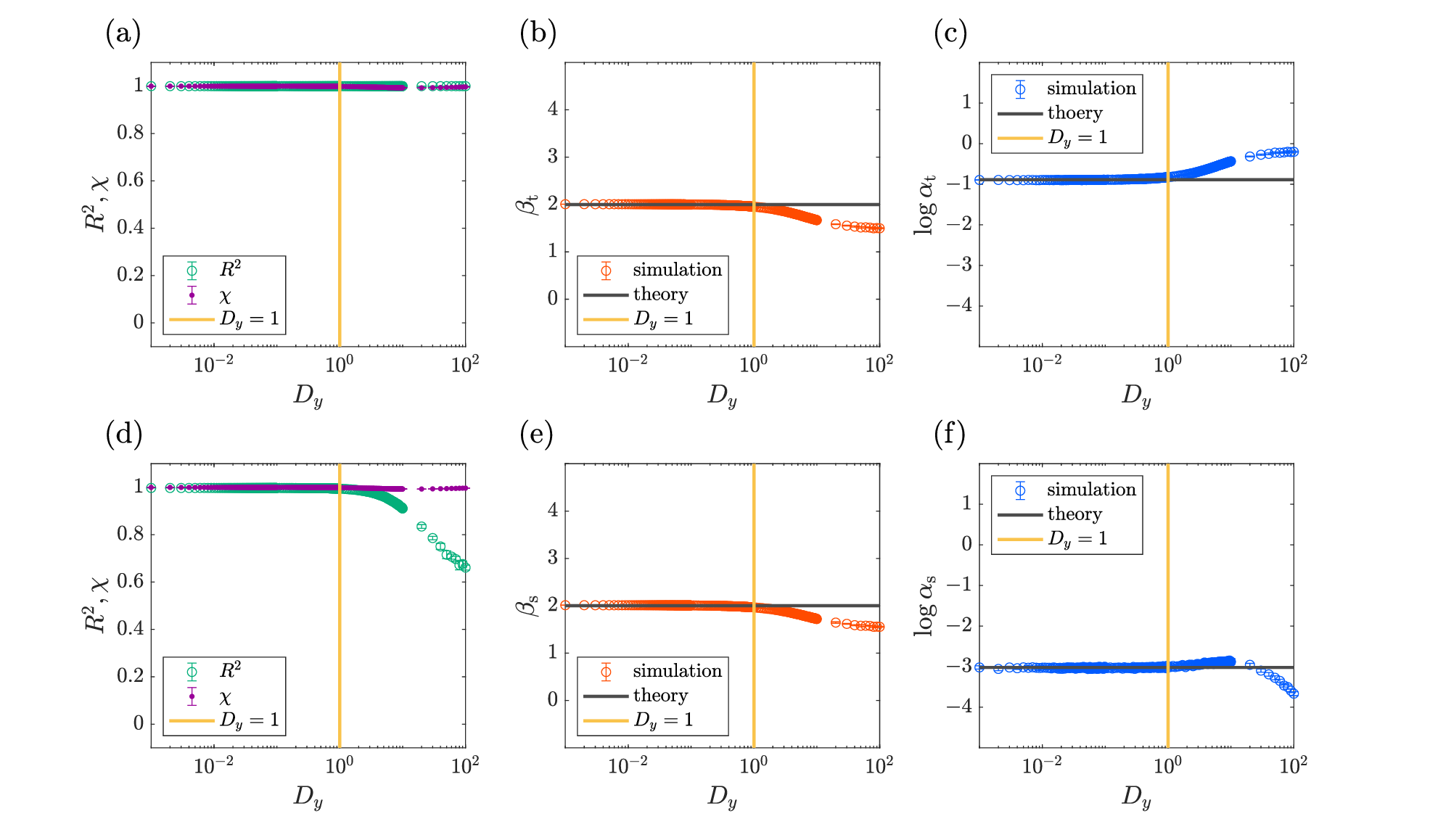}
\caption{\label{fig:hyperchaotic_Lorenz_Dy_vary} Dependence of TL parameters on $D_y$ in the globally coupled hyperchaotic Lorenz system. $a = 1, c = 0, k = 1, l = 1$, $D_x = 1000$, and $D_z = D_w = 0$. $N=100, \sigma = 10$, $b = 8/3$, and $\rho_i$ is randomly selected from a uniform distribution between $26.5$ and $29.5$. (a) $R^2$ for
temporal TL. (b) Exponents of temporal TL. (c) Intercepts of temporal TL. (d) $R^2$ for spatial
TL. (e) Exponents of spatial TL. (f) Intercepts of spatial TL.}
\end{figure*}

\begin{figure*}[hb]
\includegraphics[width=180mm]{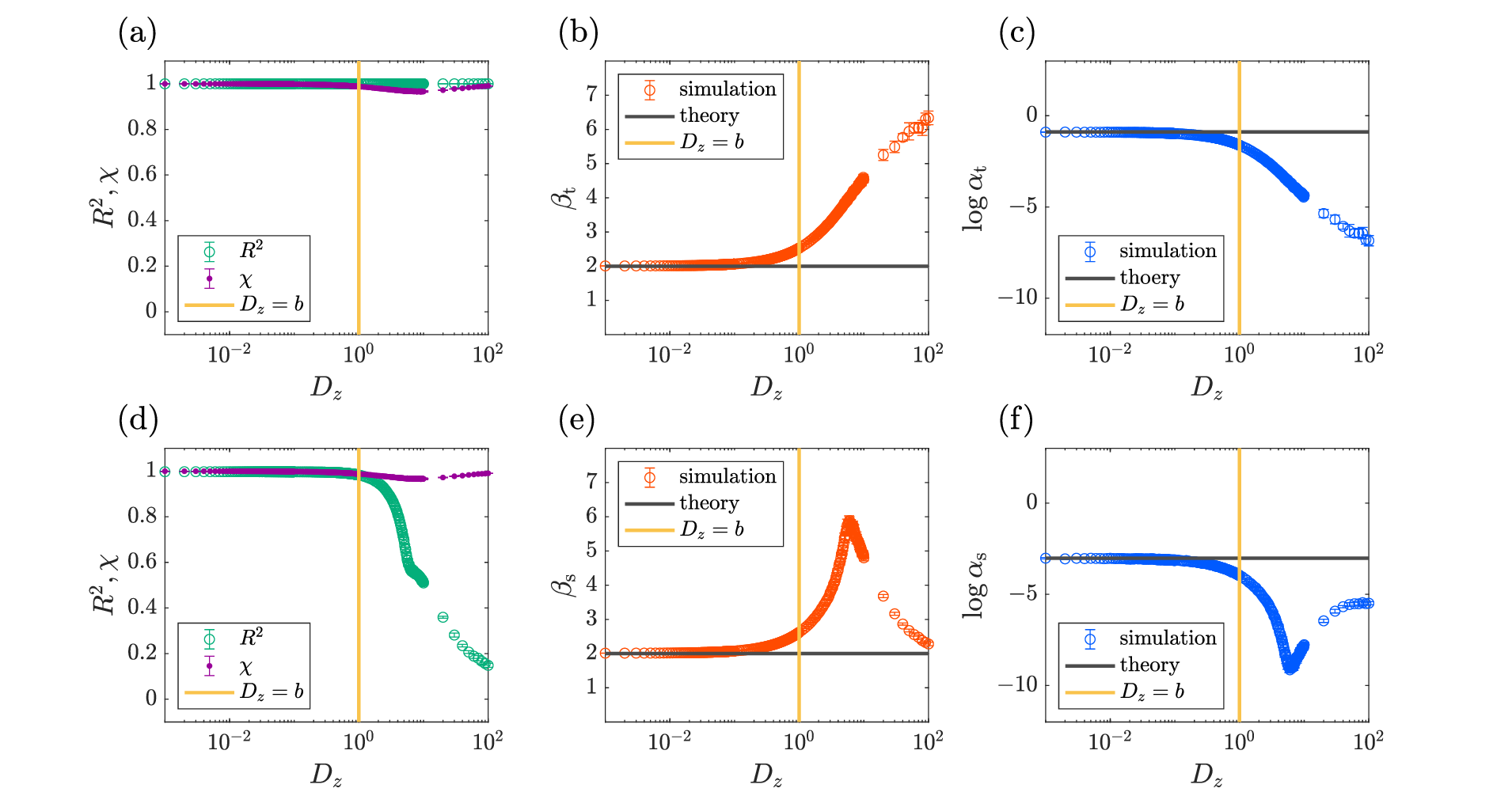}
\caption{\label{fig:hyperchaotic_Lorenz_Dz_vary} Dependence of TL parameters on $D_z$ in the globally coupled hyperchaotic Lorenz system. $a = 1, c = 0, k = 1, l = 1$, $D_x = 1000$, and $D_y = D_w = 0$. $N=100, \sigma = 10$, $b = 8/3$, and $\rho_i$ is randomly selected from a uniform distribution between $26.5$ and $29.5$. (a) $R^2$ for
temporal TL. (b) Exponents of temporal TL. (c) Intercepts of temporal TL. (d) $R^2$ for spatial
TL. (e) Exponents of spatial TL. (f) Intercepts of spatial TL.}
\end{figure*}

\begin{figure*}[hb]
\includegraphics[width=180mm]{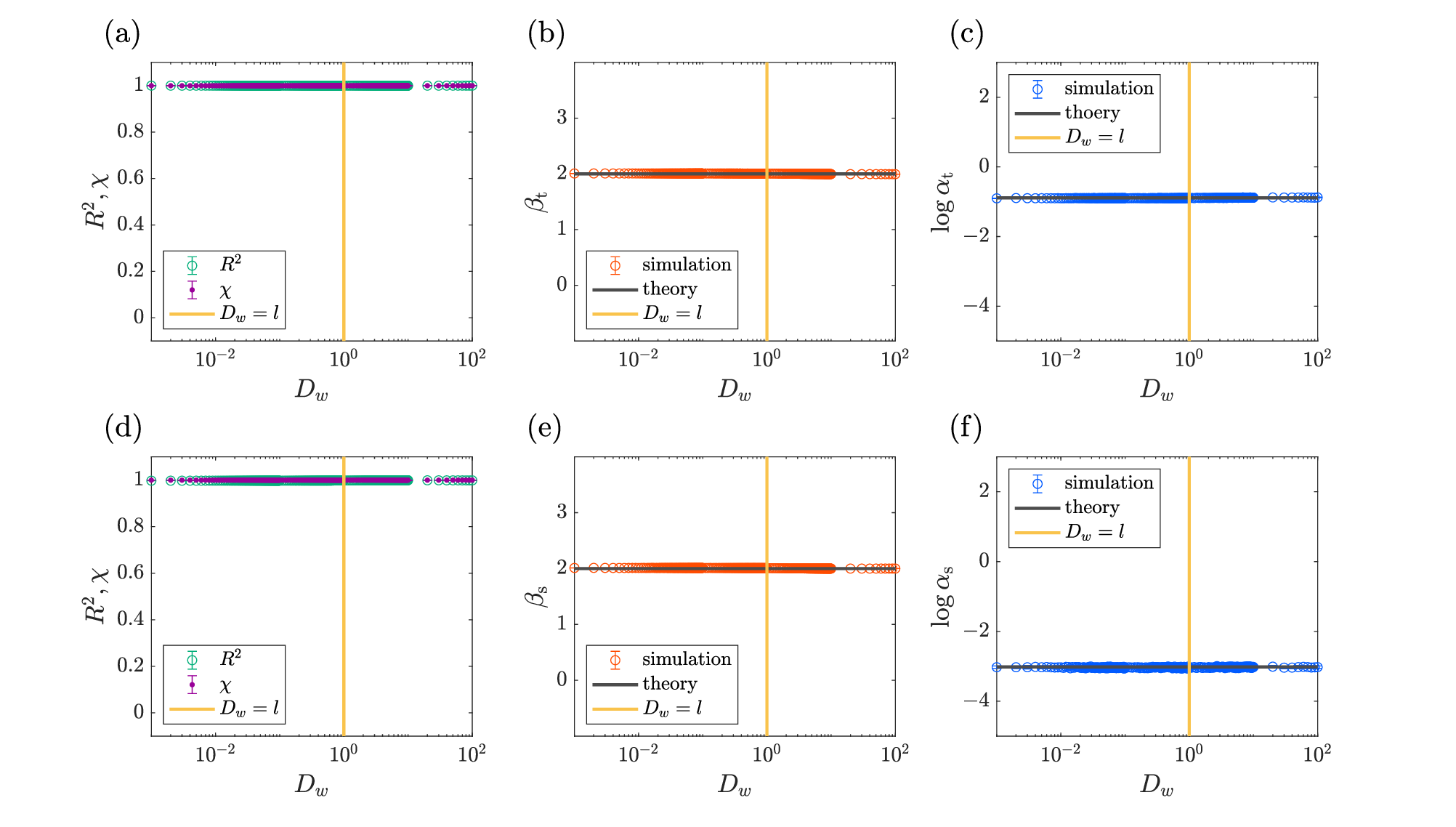}
\caption{\label{fig:hyperchaotic_Lorenz_Dw_vary} Dependence of TL parameters on $D_w$ in the globally coupled hyperchaotic Lorenz system. $a = 1, c = 0, k = 1, l = 1$, $D_x = 1000$, and $D_y = D_z = 0$. $N=100, \sigma = 10$, $b = 8/3$, and $\rho_i$ is randomly selected from a uniform distribution between $26.5$ and $29.5$. (a) $R^2$ for
temporal TL. (b) Exponents of temporal TL. (c) Intercepts of temporal TL. (d) $R^2$ for spatial
TL. (e) Exponents of spatial TL. (f) Intercepts of spatial TL.}
\end{figure*}

Next, we investigate the effects of the other coupling coefficients, $D_y$, $D_z$, and $D_w$, on the validity of the theory. Based on the analytical derivation, we expect the theoretical framework to become invalid when $D_y$ or $D_z$ is non-negligible. In contrast, our theory predicts that increasing $D_w$ does not affect the results. To test these expectations, we perform numerical simulations using the coupled hyperchaotic Lorenz system (Figs.~\ref{fig:hyperchaotic_Lorenz_Dy_vary}, \ref{fig:hyperchaotic_Lorenz_Dz_vary}, and \ref{fig:hyperchaotic_Lorenz_Dw_vary}). As shown in Fig.~\ref{fig:hyperchaotic_Lorenz_Dy_vary}, when $D_y$ becomes non-negligible compared with $1$, although $R^2$ for temporal TL remains close to 1, $\beta_{\rm t}$ and $\log \alpha_{\rm t}$ begin to deviate from the theoretical predictions, indicating a breakdown of WP. For spatial TL, when $D_y$ is non-negligible relative to $1$, $R^2$ falls below 1, and both $\beta_{\rm s}$ and $\log \alpha_{\rm s}$ also deviate from the theoretical predictions, further confirming that WP no longer holds. We can confirm that qualitatively the same behavior is observed when $D_z$ becomes non-negligible compared with $b$ (Fig.~\ref{fig:hyperchaotic_Lorenz_Dz_vary}). Furthermore, as expected, we confirm that increasing $D_w$ does not affect the emergence of TL with an exponent 2 (Fig.~\ref{fig:hyperchaotic_Lorenz_Dw_vary}).

\section{Conclusions and Discussion}
In the present paper, building upon our previous work~\cite{Mitsui_Kori_2025_PRL}, we investigated the emergence of WP and the associated TL with an exponent 2 in networks of Lorenz systems. In particular, we extended the results obtained in Ref.~\onlinecite{Mitsui_Kori_2025_PRL} by considering the globally coupled generalized and hyperchaotic Lorenz systems. As a result, we showed that WP and TL with an exponent 2 emerge in the globally coupled original, generalized, and hyperchaotic Lorenz systems in the regime of large $D_x$ and small $D_y$ and $D_z$. This indicates that the mechanism underlying these phenomena is already inherent in the original Lorenz system and persists in the generalized and hyperchaotic Lorenz systems. Theoretical predictions were in excellent agreement with numerical simulations. Moreover, the analytical calculations indicate that increasing $D_y$ or $D_z$ causes WP to break down, and consequently, TL with an exponent 2 no longer holds. Since this behavior is common to the original, generalized, and hyperchaotic Lorenz systems, it may also be viewed as an inherent property of the original Lorenz system. In contrast, our analysis also reveals that the emergence of WP and TL with an exponent 2 is insensitive to the value of $D_w$. Unlike the effects of $D_y$ and $D_z$, this property becomes apparent only through the analysis of the generalized and hyperchaotic Lorenz systems.

In the case of the four-variable model, we numerically observed that $|aw_0|$ is not always large and our assumption of a large $|aw_0|$ in our perturbative approach might be violated. Nevertheless, the numerical simulation results are in good agreement with the theoretical predictions. This observation suggests that the treatment of the $aw_0$ term in our perturbative approach requires further consideration.

Furthermore, we investigated the emergence of WP and TL with an exponent 2 in network systems. As a result, we derived the condition \eqref{eq:condition_network} which indicates that the emergence of WP and TL with an exponent 2 requires random arrangements of heterogeneous oscillators within the network. We confirmed this theoretical prediction by numerical simulations using several representative network structures such as a 2D lattice, random networks, and scale-free networks (Figs.~\ref{fig:Lorenz_TL_2dim_lattice}, \ref{fig:Lorenz_TL_random_network}, and \ref{fig:Lorenz_TL_scalefree_network}). As expected, we were able to observe WP and TL with an exponent 2 on these networks in the strong-coupling regime. In addition, we numerically investigated what happens in the case where the condition \eqref{eq:condition_network} does not hold and instead the condition \eqref{eq:condition_network_breakdown} is established. To this end, we constructed a 1D lattice system where the oscillators were arranged in order of their heterogeneity parameter values. In this case, WP and TL with an exponent 2 are not observed even under strong coupling, and instead, oscillation quenching occurs (Fig.~\ref{fig:ordered_1dim_lattice}). This result indicates that strong coupling alone does not guarantee the emergence of WP, and that the way oscillator heterogeneity is embedded in the network is important. Therefore, our results suggest that, in addition to network topology, the arrangement of oscillators is a key factor in determining whether WP emerges or oscillation quenching occurs. This suggests the possibility of controlling the collective dynamics simply by rearranging the oscillators, even without changing either the network structure or the coupling strength.
 
In this study, we chose $D_x$ as the primary coupling parameter and investigated its influence on the emergence of WP and TL. How WP and TL emerge when other coupling coefficients are taken as the primary coupling parameter will be addressed in future work.

By combining this research with the previous work~\cite{Mitsui_Kori_2025_PRL}, we provide broader evidence supporting the generality of WP and TL with an exponent 2 in strongly coupled systems. Although several mechanisms have been proposed in the literature to explain the emergence of TL with an exponent 2 (Refs.~\onlinecite{Cohen_etal_2013_PRSB,Cohen_2014_TE,Giometto_etal_2015_PNAS,Cohen_2013_TPB,Sassi_etal_2022_PRX,Kilpatrick_Ives_2003_Nature,Anderson_etal_1982_Nature,Salahshour_2023_NJP,Ballantyne_2005_EER,Carpenter_etal_2023_TPB}), the present paper and previous study~\cite{Mitsui_Kori_2025_PRL} suggest that synchronization, a universal phenomenon~\cite{synchronization_book1}, underlies TL, another universal phenomenon.

The phenomenon in which time series are proportional to each other has previously been reported using a coupled Lorenz system~\cite{Mainieri_Rehacek_1999_PRL}. In this work, we have shown that this type of synchronization appears not only in the original Lorenz system but also in the generalized and hyperchaotic Lorenz systems, thereby extending the scope of this phenomenon.

\begin{acknowledgments}
The authors thank Daiki Nishiguchi for the helpful discussions
on the relationship between Taylor's law and giant fluctuations. This study was supported by JSPS KAKENHI Grant Number 23KJ0830 and WINGS CFS program of The University of Tokyo to Y. M., and JSPS KAKENHI Grant Number 25K15258 to H. K.
\end{acknowledgments}

\section*{Author declarations}
\subsection*{Conflict of Interest}
The authors have no conflicts to disclose.

\subsection*{Author Contributions}
\noindent
\textbf{Yuzuru Mitsui:} Conceptualization, Data curation, Formal analysis, Funding acquisition (equal), Investigation, Methodology, Validation (lead), Visualization, Writing – original draft\\
\textbf{Hiroshi Kori:} Funding acquisition (equal), Supervision, Validation (supporting), Writing – review \& editing

\section*{Data Availability}
The data that support the findings of this study are available from the corresponding author upon reasonable request.

\clearpage
\appendix

\section{Analytical derivation for generalized and hyperchaotic Lorenz systems}

To show that WP and TL with an exponent 2 also hold for $z_i(t)$ in the generalized and hyperchaotic Lorenz systems, we approximately solve the coupled extended Lorenz system described by Eq.~(\ref{eq:coupled_extended_Lorenz}). As ansatzes, we consider the following expansion~\cite{Mitsui_Kori_2025_PRL}:
\begin{subequations}
  \label{eq:ansatz_extended}
\begin{align}
 x_i(t) &= x_0(t- \eps_i \tau) + \eps_i p(t- \eps_i \tau) + O(\hat{\eps}^2), \label{x_ansatz_extended}\\
 y_i(t) &= y_0(t- \eps_i \tau) + \eps_i q(t- \eps_i \tau) + O(\hat{\eps}^2), \label{y_ansatz_extended}\\
 z_i(t) &= z_0(t- \eps_i \tau) + \eps_i r(t- \eps_i \tau) + O(\hat{\eps}^2), \label{z_ansatz_extended}\\
 w_i(t) &= w_0(t- \eps_i \tau) + \eps_i s(t- \eps_i \tau) + O(\hat{\eps}^2), \label{w_ansatz_extended}
\end{align}
\end{subequations}
where $x_0$, $y_0$, $z_0$, and $w_0$ constitute a reference oscillator satisfying
\begin{subequations}
\label{eq:refosci_extended_Lorenz}
\begin{align}
\dot{x}_0 &= \sigma (y_0 - x_0) + aw_0, \label{eq:dot_extended_x0}\\
\dot{y}_0 &= x_0(\rho_0 - z_0) - y_0, \label{eq:dot_extended_y0}\\
\dot{z}_0 &= x_0 y_0 - b z_0, \label{eq:dot_extended_z0}\\
\dot{w}_0 &= cx_0 - ky_0z_0 - lw_0. \label{eq:dot_extended_w0}
\end{align}
\end{subequations}
When $ |\sigma(y_0-x_0)|, |aw_0| \gg |\mu_ip|, |\eps_i \sigma q|, |\mu_i\tau\sigma(y_0-x_0)|, |\mu_i\tau aw_0|$ is satisfied and $D_x$ is sufficiently large, the following equations can be derived by applying the same procedure in the case of the original Lorenz system.
\begin{subequations}
\label{eq:p_and_dot_qrs3}
\begin{align}
    p =& \dfrac{\sigma q + as +  D_x\tau\sigma(y_0 - x_0) + D_x\tau a w_0}{\sigma + D_x}, \label{eq:4variable_p} \\
    \dot{q} =& - q + (D_x - r)x_0, \label{eq:dot_q_later} \\
    \dot{r} =& - b r + q x_0, \label{eq:dot_r_later}\\
    \dot{s} =& - (l+D_w) s - kry_0 - kqz_0. \label{eq:dot_s}
\end{align}
\end{subequations}
Again, from Eqs.~(\ref{eq:dot_extended_y0}), (\ref{eq:dot_extended_z0}), (\ref{eq:dot_q_later}), and (\ref{eq:dot_r_later}), the proportional relationships shown in Eqs.~\eqref{eq:q_propto_y0} and \eqref{eq:r_propto_z0} are obtained. As in the original Lorenz system, the theoretical predictions for the intercept and exponent values of TL in the coupled extended Lorenz system are given by Eq.~\eqref{eq:TL_theory}.

To examine whether WP can also hold for the variable $w_i(t)$, we next consider an ansatz of the form
\begin{eqnarray}
    s(t) = \lambda w_0(t). \label{eq:w_ansatz}
\end{eqnarray}
Substituting Eq.~\eqref{eq:w_ansatz} into Eq.~\eqref{eq:dot_s}, and using Eqs.~\eqref{eq:dot_extended_w0}, \eqref{eq:q_propto_y0}, and \eqref{eq:r_propto_z0}, we obtain
\begin{align}
 \lambda cx_0 - \lambda ky_0z_0 -\lambda lw_0   = -\lambda(l+D_w)w_0 -\dfrac{2D_xk}{\rho_0}y_0z_0.
\end{align}
Therefore, when $c=0, D_w = 0$, the following relationship holds true:
\begin{eqnarray}
\label{eq:s_propto_w0}
    s(t) = \dfrac{2D_x}{\rho_0}w_0(t).
\end{eqnarray}
In other words, in the case of the coupled hyperchaotic Lorenz system in which there is no coupling in $w_i(t)$ or $D_w \ll l$, WP holds for $w_i(t)$. In this case, we can derive the quadratic relationships between the mean and variance of $
w_i(t)$ as follows:
\begin{align}
\mathrm{V}[w_i(t)]_t &= \dfrac{\mathrm{V}[w_0(t)]_t}{\mathrm{E}[w_0(t)]_t^{\, 2}}\, \mathrm{E}[w_i(t)]_t^{\, 2},  \\
\mathrm{V}[w_i(t)]_i &= \dfrac{4\mathrm{V}[\mu_i]_i}{\rho_0^{\, 2}}\,\mathrm{E}[w_i(t)]_i^{\, 2}. 
\end{align}
However, the temporal mean of $w_0(t)$ approaches $0$ in numerical simulations, making it difficult to numerically verify the emergence of temporal TL in $w_i(t)$. \\

\section{Ordered one-dimensional lattice system}
\begin{figure*}[t]
\includegraphics[width=180mm]{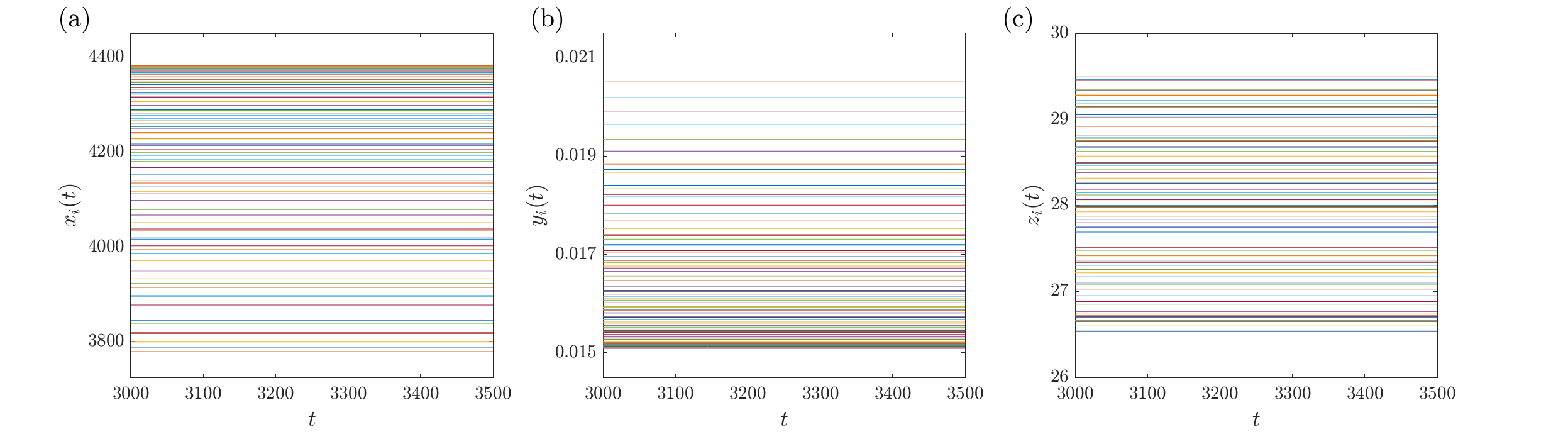}
\caption{\label{fig:ordered_1dim_lattice} Time series of the coupled Lorenz system on a 1D lattice where $\rho_1 < \rho_2 < \cdots < \rho_{100}$. $N=100, D_x=1000, \sigma = 10$, $b = 8/3$. $\rho_i$ is randomly selected from a uniform distribution between $26.5$ and $29.5$, and then arranged in ascending order. (a) Time series of $x_i(t)$. (b) Time series of $y_i(t)$. (c) Time series of $z_i(t)$.}
\end{figure*}

We provide an example where the condition \eqref{eq:condition_network} does not hold and instead the condition \eqref{eq:condition_network_breakdown} is established. Here, we consider the coupled Lorenz system on a 1D lattice where the oscillators are arranged such that the heterogeneity parameter values are in ascending order, $\rho_1 < \rho_2 < \cdots < \rho_N$. In this case, even with strong coupling, WP does not emerge and oscillation quenching occurs instead (Fig.~\ref{fig:ordered_1dim_lattice}).


%
%

%


\clearpage
\bibliography{ref}

\end{document}